\documentclass[a4paper,10pt]{article}
\usepackage[utf8]{inputenc}
\usepackage{amsmath}
\usepackage{authblk}
\usepackage{chngcntr}
\usepackage{graphicx}
\usepackage{subfig}
\usepackage{textcomp}
\usepackage{booktabs}
\usepackage{ulem}
\usepackage{amsmath}
\usepackage{amssymb}
\usepackage{color}
\usepackage{framed}
\usepackage{txfonts}
\usepackage{upgreek}
\usepackage{times}
\usepackage{bm}
\usepackage{mathbbol}
\usepackage{tensor}
\usepackage{empheq}
\usepackage[makeroom]{cancel}
\usepackage{needspace}
\usepackage{xspace}
\usepackage{relsize}
\usepackage{verbatim}
\usepackage{float}
\usepackage{bigints}
\usepackage{xcolor}
\usepackage{xfrac}
\usepackage[nice]{nicefrac}
\begin{document}
\definecolor{shadecolor}{rgb}{0.93,0.93,0.93}
\definecolor{darkred}{rgb}{0.64,0.34,0.37}
\definecolor{dunkelgrau}{rgb}{0.33,0.33,0.33}
\definecolor{eqemphcolor}{rgb}{1,1,1} 
\definecolor{eqemphcolor}{rgb}{0.93,0.93,0.93} 
\definecolor{eqemphcolor}{rgb}{1,1,0.75} 
\empheqset{box={\fboxsep=2pt\colorbox{eqemphcolor}}}
\bibliographystyle{plain}    

%
%
%
\newcommand{\RB}{\mathbb{R}}
\newcommand{\MB}{\mathbb{M}}
\newcommand{\NB}{\mathbb{N}}
\newcommand{\CB}{\mathbb{C}}
\renewcommand{\MB}{\mathbb{M}}
\newcommand{\AB}{\mathbb{A}}
\newcommand{\BB}{\mathbb{B}}
\newcommand{\im}{\mathrm{i}}
\newcommand{\FC}{F}
\newcommand{\e}{\mathrm{e}}
\newcommand{\Lag}{L}
\newcommand{\Lae}{\Lag_{\e}}
\newcommand{\HCv}{E}
\newcommand{\Hf}{H}
\newcommand{\He}{\Hf_{\e}}
\newcommand{\HCp}{\Hf^{\prime}}
\newcommand{\HCe}{\He^{\prime}}
\newcommand{\Hv}{e}
\newcommand{\HD}{\mathcal{\Hf}}
\newcommand{\HCC}{\Hf^{\prime\prime}}
\newcommand{\HCd}{\mathcal{\Hf}}
\newcommand{\FCd}{\mathcal{\FC}}
\newcommand{\LCd}{\mathcal{\Lag}}
\newcommand{\LTd}{\mathcal{T}}
\newcommand{\KCd}{\mathcal{K}}
\newcommand{\QCd}{\mathcal{Q}}
\newcommand{\PCd}{\mathcal{P}}
\newcommand{\qCd}{\mathcal{q}}
\newcommand{\Dderr}{\overrightarrow{\mathcal{D}}}
\newcommand{\Dderl}{\overleftarrow{\mathcal{D}}}
\newcommand{\bx}{\pmb{x}}
\newcommand{\by}{\pmb{y}}
\newcommand{\bk}{\pmb{k}}
\newcommand{\be}{\pmb{e}}
\newcommand{\bef}{\pmb{f}}
\newcommand{\bg}{\pmb{g}}
\newcommand{\bp}{\pmb{p}}
\newcommand{\bq}{\pmb{q}}
\newcommand{\ba}{\pmb{a}}
\newcommand{\bA}{\pmb{A}}
\newcommand{\bb}{\pmb{b}}
\newcommand{\bB}{\pmb{B}}
\newcommand{\bF}{\FC^{\mu}}
\newcommand{\bP}{\pmb{P}}
\newcommand{\bQ}{\pmb{Q}}
\newcommand{\bPhi}{\pmb{\Phi}}
\newcommand{\bphi}{\pmb{\varphi}}
\newcommand{\bPsi}{\pmb{\Psi}}
\newcommand{\bpsi}{\pmb{\psi}}
\newcommand{\bPi}{\pmb{\Pi}}
\newcommand{\bpi}{\pmb{\pi}}
\newcommand{\bvarphi}{\pmb{\varphi}}
\newcommand{\rmi}{\im}
\newcommand{\CO}{\mathcal{O}}
\renewcommand{\d}{\,\mathrm{d}}
\newcommand{\ds}{\d{s}}
\newcommand{\dt}{\d{t}}
\newcommand{\onehalf}{{\textstyle\frac{1}{2}}}
\newcommand{\twothird}{{\textstyle\frac{2}{3}}}
\newcommand{\onethird}{{\textstyle\frac{1}{3}}}
\newcommand{\quarter}{{\textstyle\frac{1}{4}}}
\newcommand{\threequarter}{{\textstyle\frac{3}{4}}}
\newcommand{\fourthird}{{\textstyle\frac{4}{3}}}
\newcommand{\oneeights}{{\textstyle\frac{1}{8}}}
\newcommand{\ihalf}{{\textstyle\frac{\rmi}{2}}}
\newcommand{\iquarter}{{\textstyle\frac{\rmi}{4}}}
\newcommand{\onetwelfths}{{\textstyle\frac{1}{12}}}
\newcommand{\dotx}{\dot{x}}
\newcommand{\dota}{\dot{a}}
\newcommand{\dotN}{\dot{N}}
\newcommand{\dott}{\dot{t}}
\newcommand{\ddota}{\ddot{a}}

\newcommand{\partialr}{\overrightarrow{\partial}}
\newcommand{\partiall}{\overleftarrow{\partial}}
\newcommand{\pfrac}[2]{\frac{\partial{#1}}{\partial{#2}}}
\newcommand{\pfracr}[2]{\frac{\partialr{#1}}{\partial{#2}}}
\newcommand{\pfracl}[2]{\frac{\partiall{#1}}{\partial{#2}}}
\renewcommand{\dfrac}[2]{\frac{\mathrm{d}{#1}}{\mathrm{d}{#2}}}
\newcommand{\ppfrac}[3]{\frac{\partial^{2}{#1}}{\partial{#2}\partial{#3}}}
\newcommand{\pppfrac}[4]{\frac{\partial^{3}{#1}}{\partial{#2}\partial{#3}\partial{#4}}}
\newcommand{\vecb}[1]{\pmb{#1}}
\newcommand{\exref}[1]{Example~\ref{#1}}
\newcommand{\pref}[1]{Part~\ref{#1}}
\newcommand{\Index}[1]{#1\index{#1}}
\newcommand{\detpartial}[2]{\left| \pfrac{#1}{#2} \right|}
\newcommand{\eref}[1]{Eq.~(\ref{#1})}
\newcommand{\qref}[1]{(\ref{#1})}
\newcommand{\fref}[1]{Fig.~\ref{#1}}
\newcommand{\sref}[1]{Section~\ref{#1}}
\newcommand{\tref}[1]{Table~\ref{#1}}
\newcommand{\dete}{\varepsilon}
\newcommand{\deteinv}{\frac{1}{\dete}}
\newcommand{\psibar}{\bar{\psi}}
\newcommand{\Psibar}{\bar{\Psi}}
\newcommand{\pibar}{\bar{\pi}}
\newcommand{\Pibar}{\bar{\Pi}}
\newcommand{\mtilde}{\mu}
\hyphenation{one-di-men-sio-nal con-fi-gu-ra-tion Ma-the-ma-ti-ca geo-metric parallel-epiped}
\normalem

\title{Setting free the cosmic time in Quantum Universe}
\author[1]{David~Vasak\thanks{vasak@fias.uni-frankfurt.de}} 

\affil[1]{Frankfurt Institute for Advanced Studies (FIAS), Ruth-Moufang-Strasse~1, 60438 Frankfurt am Main, Germany}
\maketitle


\begin{abstract}%
The standard Wheeler-DeWitt approach to Quantum Cosmology leads to a problematic freeze of time. 
 The reason for this is the enforcement of time re-parametrization invariance via the lapse function, which is treated at least partially as an independent dynamical field.
 We show that the correct treatment of the time re-parametrization gauge along the lines of the extended Hamiltonian formalism lifts the notorious ban on cosmic time as an evolution parameter in $3^{rd}$ quantized cosmology.
\end{abstract}




\paragraph{Introduction:} \label{sec:Intro}
Assuming that everything, including the Universe, is ``quantum'', led to promoting the dynamics of the classical Universe to a quantum mechanical problem.  
The gold standard of Quantum Cosmology is the time-less Wheeler-DeWitt (WDW) equation~\cite{Wheeler:1957mu, DeWitt1967}.
The evolution of the Universe in cosmic time disappeared in the underlying mathematical framework~\cite{misner73, Robles-Perez:2021rqt}.
The solutions of the WDW equation, the wave equations of the Universe,\footnote{Most notably the no-boundary solution~\cite{Hartle:1983ai, Hartle:2008ng} or the tunneling solution~\cite{vilenkin1988quantum, Vilenkin:2018dch, Vilenkin:2018oja}.} describe a Universe frozen in time.

\medskip
At the heart of that framework is the reduction of the ``superspace'' of possible spacetimes to a subspace motivated by the Cosmological Principle (CP).
CP postulates that on cosmological scales space and matter are isotropic and homogeneous.  
That ``mini superspace'' of the corresponding model geometries encompasses three 
types of the so called FLRW metrics, namely a hyperbolic, flat or spherical 
3D-space that is expanding with time.  
Matter, in addition, consists of independent non-interacting ideal fluids parametrized by their energy densities and equations of state.
In this ``Concordance model'' the classical dynamics of the Universe is 
described by the Friedman equations which are ordinary differential equations in the so called cosmic time,
for the evolution  of the set of dynamical parameters characterizing the geometry and the energy densities of matter. 

\medskip
The canonical quantization prescription promotes the classical Hamiltonian of the reduced spacetime-matter system to a quantum operator by replacing the classical coordinates and conjugate momenta with the corresponding quantum operators. 
The outcome is a Schr\"{o}dinger-like wave equation. 
Unfortunately, in this process the classical Hamiltonian appears to be constrained to zero, which means that its quantum version has only a zero eigenvalue and the wave function is time independent.
In consequence the quantum Universe is frozen in terms of the cosmic time. 
That finding has ever since been accepted by the cosmology community as an undisputed fact and addressed by a plethora of papers.

However, and this is the topic of this short note, the process of defining the cosmological Hamiltonian is not unambiguous. 
We show that the evolution in terms of the canonical cosmic time can be retained in Quantum Cosmology by properly identifying the physical role of the lapse function. 

\paragraph{The cosmological mini superspace:}
Upon foliation of spacetime in spacelike Cauchy hypersurfaces separated by infinitesimal time leaps, the spacetime metric (signature +2, greek indices are 4D, latin indices are 3D, natural units $\hbar = c = 1$) becomes
\begin{equation*}
 \d s^2 = g_{\mu\nu} \,\d x^\mu \d x^\nu = \left(N_a N^a - N^2 \right) \d t^2 +2N_a \, \d x^a \d t + h_{ab}\, \d x^a \d x^b.
\end{equation*} 
The spatial metric, shift vector and lapse function, $h_{ab}, N^a$ and $N$ respectively, are a small set of parameters restricting the vast superspace of possible metrics to a mini subspace that appears reasonably aligned with astronomical observations.
Usually, it is assumed a priori that it is possible to ``gauge away'' the shift vector, that is, to find local transition maps such that $N^a = 0$ holds everywhere.
In addition, the spatial geometry represented by the spatial metric $h_{ab}$ is further simplified to isotropic spaces (spherical, flat or hyperboloid) with the scale factor $a(t)$. 
The resulting mini superspace applied in the following is the FRLW metric,
\begin{equation}\label{def:FLRW}
 \d s^2 = -N^2(t) \, \d t^2 + a^2(t) \,\d \Omega_3^2.
\end{equation}
Notice that here it still includes the lapse function retaining the (gauge) freedom 
for re-parametrizing the time variable. 
In the standard approach to classical cosmology, though, $N=1$ is set. 
The metric~\eqref{def:FLRW} with $N=1$ is then inserted into Einstein's field equation yielding the Friedman and acceleration equations:
\begin{subequations}\label{eq:Friedman}
 \begin{align}
  H^2(a) &:= \left(\frac{\d a / \d t}{a}\right)^2= \frac{8\pi G}{3} \rho(a),\\
  \frac{\d^2a}{\d t^2} &= \frac{4\pi G}{3} \left[\rho(a) + 3p(a)\right].
\end{align}
\end{subequations}
$H(a)$ is called Hubble parameter.
Driven by the density and pressure functions, $\rho(a)$ and $p(a)$, encompassing radiation, (dark) matter, dark energy, and spatial curvature, the Friedman equation determines the evolution of the Universe in cosmic time $t$.
The exact form of the density and pressure functions depends on the specific gravity theory and in general contains further fields and parameters that are not essential for the problem in question.

\medskip
There is an obvious formal analogy of the Friedman equations to the dynamics of a single non-relativistic massive point particle moving in the potential $V(a)$.
In general the pressure is some function of energy density, $p = p(\rho)$, called equation of state. 
For consistency that function must be such that equations~\eqref{eq:Friedman} can be re-written as  
\begin{subequations}\label{eq:Friedman2}
 \begin{align}
 &\left(\frac{\d a}{\d t}\right)^2 + V(a) = 0 \label{eq:Friedman21}\\
 &\frac{\d^2a}{\d t^2} = -\frac{1}{2} \frac{\d V}{\d a} \label{eq:Friedman22}
\end{align}
\end{subequations}
with $V(a) := -\tfrac{8\pi G\,a^2}{3} \rho(a)$.
Secondly, with the Lagrangian defined as   
\begin{equation}
 L(a,\tfrac{\d a}{\d t},...;t) = \frac{m}{2} \left(\frac{\d a}{\d t}\right)^2 -V(a), 
\end{equation}
the acceleration equation,~\eqref{eq:Friedman22}, can be obtained by variation of the action integral 
\begin{equation*}
 S = \int \d t\, L(a,\tfrac{\d a}{\d t},...;t),
\end{equation*}
where we set $m = 2$ for the fictitious particle's (dimensionless) mass.
The first Friedman equation~\eqref{eq:Friedman21} fixes the particle's total energy to zero.


\paragraph{The standard approach and freeze of time in Quantum Cosmology:}
According to the standard approach to quantum cosmology, see e.g.~\cite{Robles-Perez:2021rqt}, the lapse function facilitates via $dt = N(s) \,ds$ an arbitrary re-parametrization of  the time coordinate.
The FLRW metric~\eqref{def:FLRW} is inserted into the Einstein-Hilbert action integral that can be brought into the form
\begin{equation}\label{action}
 S = \int \d s \,N(s) \, L(a,\dota/N,...;s) = \int \d s \,L'(a,\dota,N;s).
\end{equation}
The transformed Lagrangian $ L'(a,\dota,N;s) := N \, L(a,\dota/N,...;s)$ with $\dota := \tfrac{\d a}{\d s}$ depends on the \emph{field} $N(s)$ but not on its derivative:
\begin{equation}\label{def:Lagrangianpp}
 L'(a,\dota,N;s) = N \left(\frac{\dota^2}{N^2}-V(a)\right).
\end{equation}
The corresponding Euler-Lagrange equations,
\begin{subequations}
\begin{align}
 0 &= \frac{\d}{\d s}\pfrac{L'}{\dota} - \pfrac{L'}{a} \\
 0 &= \frac{\d}{\d s}\pfrac{L'}{\dotN} - \pfrac{L'}{N},
\end{align}
\end{subequations}
yield upon evaluation, and after setting $N=1$, 
\begin{subequations}
\begin{align}
 \ddota + \onehalf \frac{\d V}{\d a} &= 0 \label{eq:fr2mod} \\
  \dota^2 + V(a) &= 0 \label{eq:fr1mod}.
 \end{align}
\end{subequations}
This obviously reproduces Eqs.~\eqref{eq:Friedman2}!
Moreover, we also observe that \eref{eq:fr2mod} is related to the derivative of \eref{eq:fr1mod}:
\begin{equation}
2\dota\left(\ddota + \onehalf \frac{\d V}{\d a}\right) = \frac{\d}{\d s} \left( 
\dota^2 + V(a) \right) = 0.
\end{equation}
While the r.h.s. of the above equation allows for a less restrictive energy constraint, namely
\begin{equation}
 \dota^2 + V(a) = E = const. ,
\end{equation}
But $E = 0$ is enforced by~\eref{eq:fr1mod}.

\medskip
In order to express the above analysis in the Hamiltonian picture, we first identify the conjugate momenta:
\begin{subequations}\label{def:conjmom}
 \begin{align} 
p_a &:= \pfrac{L'}{\dota} = \frac{2\dota}{N} \\
p_N &:= \pfrac{L'}{\dotN} \equiv 0 .\label{def:pN}
\end{align}
\end{subequations}
The Legendre transformation yields the corresponding Hamiltonian that is independent of $p_N$, with the lapse function factored out:
\begin{equation*}
 H' = p_N \, \dotN + p_a \, \dota - L' = N\left(\quarter p_a^2 + V\right) =: N\, \HD(a,p_a),
\end{equation*}
After adding the Dirac constraint for $p_N = 0$ via the Lagrange multiplier, $\lambda$, to the Hamiltonian, 
the canonical equations give
\begin{subequations}
 \begin{align} \label{eq:canonical1}
 \pfrac{(H'+\lambda \, p_N)}{\lambda} &= p_N = 0,\\
 \dot{p}_N = -\pfrac{(H'+\lambda \, p_N)}{N} &= -\HD.
\end{align}
\end{subequations}
After $N = 1$ is set to ensure that the cosmic time is maintained in the 
Friedman equations, combining these equations recovers the result found in the Lagrangian picture.

\medskip
Quantum cosmology emerges here from the Friedman formalism by the so called $3^{rd}$ quantization\footnote{See Kucha\v r~\cite{Kuchar:1991qf} for the different quantization categories.}. 
The variables $a, p_a$ in the vanishing Hamiltonian $\HD$ are thereby replaced 
by operators $\hat{a} \equiv a, \hat{p}_a = -\im \pfrac{}{a}$, yielding the Schr\"odinger-like equivalent
\begin{equation*}
 \hat{\HD}(\hat{a},\hat{p}_a) \, \psi(a) = 0.
\end{equation*}
This equation known as the Wheeler-DeWitt equation is timeless. 
The quantum universe is frozen w.r.t. the cosmic time. 


\paragraph{The lapse function is absorbed in a field derivative:}
Here we interpret the time as a further independent coordinate function $t(s)$ of the evolution parameter $s$, such that the re-parametrization $dt = N(s) \,ds$ is substituted by 
the velocity function, 
\begin{equation*}
N(s) \rightarrow \frac{\d t}{\d s}\equiv \dott.  
\end{equation*}
The lapse function $N(s)$ is traded for the \emph{derivative} of the time field $t(s)$~\cite{Kiefer:2025udf} 
that extends the number of dynamical degrees of freedom~\cite{struckmeier09,lanczos49} 
in the system\footnote{A related idea is known from the work of Guendelman et al. \cite{Guendelman:1996qy, Guendelman:1999qt}, where the covariant measure of integration is substituted by a total derivative of dynamical fields.}.
The Lagrangian~\eqref{def:Lagrangianpp} then becomes
\begin{equation}\label{lagrangian2}
 L'(a,\dota,\dott;s) = \dott \, \left(\frac{\dota^2}{\dott^2} -V(a) \right).
\end{equation}
The Euler-Lagrange equations now read
\begin{subequations}
\begin{align}
 0 &= \frac{\d}{\d s}\pfrac{L'}{\dota} - \pfrac{L'}{a} \\
 0 &= \frac{\d}{\d s}\pfrac{L'}{\dott} - \pfrac{L'}{t}.\label{pN}
\end{align}
\end{subequations}
Again, $\dott(s) \equiv N(s) = 1$ can be stipulated a posteriori and the (cosmic) time recovered,
giving two identical equations:
\begin{subequations}
\begin{align}
 &\ddota = -\onehalf \frac{\d V}{\d a} \label{eq:fr2mod1}\\
 &\ddota = -\onehalf \frac{\d V}{\d a}.
 \end{align}
\end{subequations}
We observe that the constraint equation~\eqref{eq:fr1mod} and hence the restriction to zero total energy is missing here.

\medskip
In the Hamiltonian picture the subtle difference in handling the lapse function leads to a modification of the momenta~\eqref{def:conjmom}:
\begin{subequations}
\begin{align}
 p_a &:= \pfrac{L'}{\dota} = \frac{2\dota}{\dott}\label{pa}\\
 p_t &:= \pfrac{L'}{\dott} = -\left[ \left(\frac{\dota}{\dott}\right)^2  + V(a) \right] \label{pt}
\end{align} 
\end{subequations}
$p_t$ is here the conjugate momentum to $t$, the time variable\footnote{This denies Kiefer in Section 3 of his book~\cite{Kiefer:2025udf}, and hence overlooks the implication for the cosmic time.}. 
Moreover, the Euler-Lagrange equations w.r.t. the coordinate $t$ and the fact that $L'$ does not depend on $t$ yields:
\begin{equation}
0 =  \frac{\d}{\d s}\,\pfrac{L'}{\dott} - \pfrac{L'}{t} = \frac{\d p_t}{\d s}. 
\end{equation}
Hence $p_t = const =: -E$.
Performing the Legendre transform yields the Hamiltonian
\begin{equation*}
 H' = p_t\,\dott + p_a \, \dota - L' = \dott \left(\quarter p_a^2 + V + p_t \right) =: \dott \, \left( \HD(a,p_a) + p_t\right).
\end{equation*}
For comparison with the canonical Eqs.~\eqref{eq:canonical1} we set w.l.o.g. $\dott = N$ and attach the Dirac constraint for $p_t = -E$ to obtain:
\begin{align*}
 \pfrac{\left[N(\HD + p_t) +\lambda \, (p_t + E)\right]}{\lambda} &= p_t + E = 0 \\
 \pfrac{(\left[N(\HD + p_t) +\lambda \, (p_t + E)\right]}{N} &= \HD + p_t = 0.
\end{align*} 
Hence the resulting constraint equation\footnote{Notice that it is not an identity, see the discussion in~\cite{struckmeier09, Struckmeier:2024ljl} in the context of the extended Hamiltonian formalism.} is now
\begin{equation}
 \HD(a,p_a) - E = 0. 
\end{equation}
Looking back we see that the difference between the constraints in Eqs.~\eqref{def:pN} and~\eqref{pt} has been crucial for abandoning the condition $E \equiv 0$.
In Quantum Cosmology this means that the Wheeler-DeWitt equation is modified to 
\begin{equation}\label{eq:HamiltonianE}
 \hat{\HD}(\hat{a},\hat{p}_a,...) \, \psi(a,...) = \hat{E}\, \psi(a,...),
\end{equation}
Since  time and energy are conjugate variables, $\hat{E} := \im \pfrac{}{t}$, yielding 
the time dependent Schr\"odinger equation 
\begin{equation}\label{eq:Hamiltoniant}
 \hat{\HD}(\hat{a},\hat{p}_a) \, \psi(a,t) =  \im \pfrac{}{t}\, \psi(a,t).
\end{equation}
The cosmological time is retained as the evolution parameter.

\paragraph{Summary:}
In the process of $3^{rd}$ quantization of cosmology and derivation of the Wheeler-DeWitt equation, the time re-parametrization invariance w.r.t. the lapse function $N(s)$ causes freezing the evolution of the Quantum Universe in cosmic time.
The underlying reason is that the lapse function, which is a mere gauge parameter, is treated as an independent dynamical entity. 
In an alternative approach, along the lines of the extended Hamiltonian formalism developed by 
Struckmeier~\cite{Struckmeier:2024ljl}, that parameter is absorbed in a (gauge) coordinate $t(s)$, constituting an additional dynamical physical degree of freedom.  
This almost trivial modification  lifts the zero constraint for the classical Hamiltonian, and sets free the cosmic time in Quantum Universe.


\paragraph{Acknowledgments:}
This work has been supported by the Walter Greiner Gesellschaft zur F\"orderung der physikalischen Grundlagenforschung e.V.
DV thanks especially the Fueck-Stiftung for support.
This work benefitted from J\"urgen Struckmeier's expertise on the extended 
Hamiltonian formalism, from the critical and stimulating discussions with 
colleagues of the Canonical Gravity group at FIAS, from inspiring 
conversations with Eduardo Guendelman, and from the critical comments of two unknown referees. 

\paragraph{Conflicts of interest:} The authors declare no conflicts of interest.

\paragraph{Data availability:} There is no associated data with this article, and as such, no new data was
generated or analyzed in support of this research.

\paragraph{Declaration of competing interest:} The authors declare that they have no known competing financial interests or personal relationships that could have appeared to influence the work reported in this paper.

\paragraph{Declaration on use of generative AI:} The authors state that they have not used any AI tools as part of the study process.

\pagebreak

\begin{thebibliography}{10}

\bibitem{DeWitt1967}
Bryce~S. DeWitt.
\newblock Quantum theory of gravity. i. the canonical theory.
\newblock {\em Phys. Rev.}, 160:1113--1148, Aug 1967.

\bibitem{Guendelman:1999qt}
E.~I. Guendelman.
\newblock {Scale invariance, new inflation and decaying lambda terms}.
\newblock {\em Mod. Phys. Lett. A}, 14:1043--1052, 1999.

\bibitem{Guendelman:1996qy}
E.~I. Guendelman and A.~B. Kaganovich.
\newblock {The Principle of nongravitating vacuum energy and some of its
  consequences}.
\newblock {\em Phys. Rev. D}, 53:7020--7025, 1996.

\bibitem{Hartle:1983ai}
J.~B. Hartle and S.~W. Hawking.
\newblock {Wave Function of the Universe}.
\newblock {\em Phys. Rev. D}, 28:2960--2975, 1983.

\bibitem{Hartle:2008ng}
James~B. Hartle, S.~W. Hawking, and Thomas Hertog.
\newblock {The Classical Universes of the No-Boundary Quantum State}.
\newblock {\em Phys. Rev. D}, 77:123537, 2008.

\bibitem{Kiefer:2025udf}
Claus Kiefer.
\newblock {\em {Quantum Gravity}}.
\newblock International Series of Monographs on Physics. Oxford University
  Press, 5 2025.

\bibitem{Kuchar:1991qf}
K.~V. Kuchar.
\newblock {Time and interpretations of quantum gravity}.
\newblock {\em Int. J. Mod. Phys. D}, 20:3--86, 2011.

\bibitem{lanczos49}
C.~Lanczos.
\newblock {\em {The Variational Principles of Mechanics}}.
\newblock University of Toronto Press, Toronto, Ontario Reprint 4th edn (Dover
  Publications, New York, 1986), 1949.

\bibitem{misner73}
Charles~W. Misner, K.~S. Thorne, and J.~A. Wheeler.
\newblock {\em {Gravitation}}.
\newblock W. H. Freeman, San Francisco, 1973.

\bibitem{Robles-Perez:2021rqt}
Salvador~J. Robles-P\'erez.
\newblock {Quantum Cosmology with Third Quantisation}.
\newblock {\em Universe}, 7(11):404, 2021.

\bibitem{struckmeier09}
J.~Struckmeier.
\newblock {Extended {H}amilton-{L}agrange formalism and its application to
  {F}eynman's path integral for relativistic quantum physics}.
\newblock {\em Int. J. Mod. Phys. E}, 18:79, 2009.

\bibitem{Struckmeier:2024ljl}
J\"urgen Struckmeier and Walter Greiner.
\newblock {\em {Extended Lagrange and Hamilton Formalism for Point Mechanics
  and Covariant Hamilton Field Theory}}.
\newblock World Scientific, 9 2024.

\bibitem{vilenkin1988quantum}
Alexander Vilenkin.
\newblock Quantum cosmology and the initial state of the universe.
\newblock {\em Physical Review D}, 37(4):888, 1988.

\bibitem{Vilenkin:2018dch}
Alexander Vilenkin and Masaki Yamada.
\newblock {Tunneling wave function of the universe}.
\newblock {\em Phys. Rev. D}, 98(6):066003, 2018.

\bibitem{Vilenkin:2018oja}
Alexander Vilenkin and Masaki Yamada.
\newblock {Tunneling wave function of the universe II: the backreaction
  problem}.
\newblock {\em Phys. Rev. D}, 99(6):066010, 2019.

\bibitem{Wheeler:1957mu}
John~A. Wheeler.
\newblock {On the Nature of quantum geometrodynamics}.
\newblock {\em Annals Phys.}, 2:604--614, 1957.

\end{thebibliography}

\end{document}